\documentclass[aps,prl,showpacs,amssymb]{revtex4}

\usepackage{graphicx}
\usepackage{bm}

\begin{document}
\title{Spinning strings, cosmic dislocations, and chronology protection}
\author{V. A. De Lorenci}
 \email{delorenci@unifei.edu.br}
\author{E. S. Moreira, Jr.}
 \email{moreira@unifei.edu.br}
\affiliation{Instituto de Ci\^encias Exatas,
Universidade Federal de Itajub\'a,
Av.\ BPS 1303 Pinheirinho, 37500-903 Itajub\'a, Minas Gerais, Brazil}

\date{September, 2003}

\begin{abstract}
A massless scalar field is quantized in the 
background of a spinning string with cosmic dislocation.
By increasing the  spin density toward the dislocation
parameter, a region containing closed 
timelike curves (CTCs) eventually forms
around the defect. Correspondingly, the propagator tends to  the
ordinary cosmic string propagator, leading therefore
to a mean-square field fluctuation, which remains well behaved
throughout the process, unlike
the vacuum expectation value of the energy-momentum tensor, which
diverges due to a subtle mechanism.
These results suggest that back reaction leads to the formation of
 a ``horizon" that protects from the appearance of CTCs.

\end{abstract}
\maketitle

%
%
Investigations on quantum theory around spinning defects go back to the 
late 1980s with the study of quantum mechanics of relativistic particles
on the spinning cone \cite{ger89}. Such a background is the Kerr-like solution
of the Einstein equations in three dimensions, whose line element is given
by (throughout the text $c=\hbar=1$, and the metric parameters are nonnegative)
\begin{equation}
ds^{2}=(d\tau+Sd\theta)^{2}-dr^{2}-\alpha^{2}r^{2}d\theta^{2},
\label{3le}
\end{equation}
where $S$ and $\alpha$ are the spin and the disclination parameter, respectively
\cite{des84}. 
Clearly Minkowski spacetime corresponds to  $S=0$ and $\alpha=1$.
Lifting the geometry  in Eq. (\ref{3le}) to four dimensions, one
obtains the gravitational background around a spinning cosmic string \cite{maz86},
for which
\begin{equation}
ds^{2}=(d\tau+Sd\theta)^{2}-dr^{2}-\alpha^{2}r^{2}d\theta^{2}-d\xi^{2}.
\label{4le}
\end{equation}

An inspection of Eqs. (\ref{3le}) and (\ref{4le}) shows that
the region for which $r<S/\alpha$ contains CTCs,
resulting that when $S\neq 0$ the corresponding spacetimes are not globally hyperbolic.
It is not clear if quantum theory makes sense in nonglobaly hyperbolic spacetimes \cite{ful89}.
In fact, quantum mechanics on the spinning cone has shown that  
$S\neq 0$ spoils unitarity \cite{ger89}.
In the context of the second quantization around spinning cosmic strings
\cite{mat90}, a recent analysis has revealed that
a nonvanishing spin density $S$ leads to divergent vacuum fluctuations \cite{lor01}.

In order to recover boost invariance along the symmetry axis,
the authors in Ref.  \cite{gal93} have ``amended" the geometry in
Eq. (\ref{4le}) by postulating a cosmic dislocation, such that
\begin{equation}
ds^{2}=(d\tau+Sd\theta)^{2}-dr^{2}-\alpha^{2}r^{2}d\theta^{2}
-(d\xi+\kappa d\theta)^{2},
\label{5le}
\end{equation}
whose metric tensor fits as solution of the Einstein equations, as well as solution
of the Einstein-Cartan equations \cite{tod94,let95}.
When $S>\kappa$, the region for which 
$r<\sqrt{S^{2}-\kappa^{2}}/\alpha$
contains CTCs. When $S<\kappa$ though, 
the spacetime is globally hyperbolic.

Vacuum fluctuations typically diverge on the Cauchy surface
(chronology horizon), which separates a region with CTCs from another
without CTCs (for a review see Ref. \cite{mat02}).
This fact has led to the 
chronology protection conjecture, 
according to which, physical laws do not allow the appearance
of CTCs (``time machines'') \cite{haw92}.
Although the geometry in  Eq. (\ref{5le}) does not contain any
Cauchy horizon [for $S>\kappa$, Eq. (\ref{5le})
describes an ``eternal time machine''], it might be clarifying  
to study quantum effects in the corresponding spacetime as the metric parameters 
are adjusted such that  CTCs are about to form. Using a massless scalar field
as a probe,
this work implements such an investigation by considering $S<\kappa$ and by taking 
$S\rightarrow\kappa$, i.e., arbitrarily close
to the point when the spacetime is about to become nonglobally 
hyperbolic.

According to Ref. \cite{gal93}, 
when the metric parameters in Eq. (\ref{5le}) satisfy
$S<\kappa$, a Lorentz frame exists with respect
to which the spin density vanishes 
(and one might say, in this case, that $S\neq 0$ is a kinematic effect). Indeed,
by performing the following Lorentz transformation in the $\tau-\xi$ plane,
\begin{eqnarray}
&&t=\frac{\tau-v\xi}{\sqrt{1-v^{2}}}\hspace{1.0cm}
z=\frac{\xi-v\tau}{\sqrt{1-v^{2}}}\hspace{1.0cm}v:=S/\kappa,
\label{boost}
\end{eqnarray} 
Eq. (\ref{5le}) can be recast as
\begin{equation}
ds^{2}=dt^{2}-dr^{2}-\alpha^{2}r^{2}d\theta^{2}
-(dz+\kappa' d\theta)^{2},
\label{dle}
\end{equation}
describing the gravitational background of a cosmic dislocation
with dislocation parameter 
\begin{equation}
\kappa':=\sqrt{\kappa^{2}-S^{2}},
\label{dparameter}
\end{equation}
and for which the usual identification
\begin{equation}
(t,r,\theta,z)\sim (t,r,\theta+2\pi,z)
\label{identification1}
\end{equation}
is observed. It should be noted that the background of 
a spinning string can only be seen as that of a cosmic dislocation 
when $S<\kappa$, 
since when $S\geq\kappa$ the boost in Eq. (\ref{boost}) becomes singular.

Vacuum fluctuations of a massless scalar field $\phi$ around a 
cosmic dislocation have recently been reported in the literature \cite{lor03}.
Equations (\ref{dle}) and (\ref{identification1}) show that  when 
$\kappa'\rightarrow 0$
the corresponding vacuum fluctuations
become those  associated with an ordinary 
cosmic string (see, e.g., \cite{smi90}). 
It follows that all scalar vacuum averages observed from the frame
corresponding to Eq. (\ref{5le})  are meant to remain finite
as $S$ is taken arbitrarily close to the  critical value $\kappa$.
For example, as  
$S\rightarrow\kappa$
the mean-square field fluctuation approaches   
\begin{equation}
\langle\phi ^{2}(r)\rangle=
\frac{1}{48\pi^{2}r^{2}}
\left(\alpha^{-2}-1\right),
\label{dphi2}
\end{equation}
which is finite (away from the defect).

Turning to the vacuum expectation value of the energy-momentum tensor
$\left<{\cal T}^{\mu}{}_{\nu}\right>$,
it is more convenient to
use local inertial coordinates 
$({\rm T},r,\varphi,\Xi)$ and  $(t,r,\varphi,Z)$ 
associated with Eqs. (\ref{5le}) and  (\ref{dle}), respectively, which are defined as
${\rm T}:= \tau + S\theta$, $\varphi:=\alpha\theta$, 
$\Xi:=\xi+ \kappa\theta$ and $Z:=z+\kappa'\theta$.
In terms of  these coordinates, both
Eqs. (\ref{5le}) and  (\ref{dle}) become the Minkowski line
element written in cylindrical coordinates, and Eq. (\ref{identification1})
leads to
$(t,r,\varphi,Z)\sim (t,r,\varphi+2\pi\alpha,Z+2\pi\kappa')$,
revealing  a ``space-helical'' structure. 
[Representing the spinning string by a rotating helix,
Eq. (\ref{boost}) leads to  the frame travelling
through the symmetry axis, and for which the helix does not rotate].
It should be mentioned that 
when the coordinates $({\rm T},r,\varphi,\Xi)$
are used, the background appears to have also a ``time-helical'' structure \cite{des84}.

Using Eq. (\ref{boost}), 
one finds out that the energy density 
$\left<{\cal T}^{{\rm T}}{}_{{\rm T}}\right>$ 
in the spinning string inertial frame
$({\rm T},r,\varphi,\Xi)$ 
is related with
$\left<T^{\mu}{}_{\nu}\right>$ 
in the cosmic dislocation inertial frame 
$(t,r,\varphi,Z)$ 
by
\begin{equation}
\left<{\cal T}^{{\rm T}}{}_{{\rm T}}\right>=
\frac{\left<T^{t}{}_{t}\right>-v^{2}\left<T^{Z}{}_{Z}\right>}
{1-v^2}.
\label{edensity}
\end{equation}
A superficial investigation 
may suggest that the relativistic factor in Eq. (\ref{edensity})
will make $\left<{\cal T}^{{\rm T}}{}_{{\rm T}}\right>$ to diverge
as $S\rightarrow\kappa$
[$v\rightarrow 1$, cf. Eq. (\ref{boost})]. 
However, as will be seen shortly, this is incorrect --- 
$\left<{\cal T}^{{\rm T}}{}_{{\rm T}}\right>$ 
indeed diverges as $S\rightarrow\kappa$; but the mechanism through which that operates 
is rather subtle and does not involve any relativistic factor.
Another pitfall consists of carrying over Eq. (\ref{edensity})
the fact that $S\rightarrow\kappa$ 
[$\kappa'\rightarrow 0$, cf. Eq. (\ref{dparameter})] leads to
$\left<T^{Z}{}_{Z}\right>\rightarrow\left<T^{t}{}_{t}\right>$
\cite{lor03}, and then to conclude (incorrectly) that 
$\left<{\cal T}^{{\rm T}}{}_{{\rm T}}\right>\rightarrow\left<T^{t}{}_{t}\right>$.
The flaw in this argument will be clear in the following.

At this point one recalls that 
$\left<T^{t}{}_{t}\right>$ and $\left<T^{Z}{}_{Z}\right>$
in Eq. (\ref{edensity}) are vacuum fluctuations in the
background of a cosmic dislocation with dislocation
parameter $\kappa'$. As is explained in Ref. \cite{lor03}, 
$\left<T^{t}{}_{t}\right>$ and $\left<T^{Z}{}_{Z}\right>$
can be  obtained by letting a certain differential operator 
to act on the corresponding renormalized propagator 
$D^{(\alpha , \kappa')}(x,\bar{x})$, according to the prescription in 
Eq. (17) of Ref. \cite{lor03}.
Proceeding along these lines, it follows that
\begin{equation}
\left<T^{Z}{}_{Z}\right> - \left<T^{t}{}_{t}\right> = 
-i\lim_{\bar{x}\rightarrow x}\left(\partial_{t}\partial_{\bar{t}} 
+ \partial_{Z}\partial_{\bar{Z}}\right)D^{(\alpha , \kappa')}(x,\bar{x}),
\label{Zt}
\end{equation} 
and by  letting the derivatives to act on the expression  of 
$D^{(\alpha , \kappa')}(x,\bar{x})$ in Eq. (12) of Ref. \cite{lor03},
Eq. (\ref{Zt}) yields
\begin{equation}
\left<T^{Z}{}_{Z}\right>=\left<T^{t}{}_{t}\right>+\frac{\kappa'^2}{r^6}
f_{\alpha}\left(\kappa'^2/r^2\right),
\label{drelation}
\end{equation}
where
\begin{equation}
f_\alpha(x):=
-\frac{1}{2}\int^{\infty}_{0}d\lambda\sum_{n=1}^{\infty}
\frac{n^2\left[\lambda^2-\pi^2(4\alpha^2 n^2-1)\right]}
{\left[\pi^2(2\alpha n+ 1)^2
+ \lambda^2\right] \left[\pi^2(2\alpha n - 1)^2 +\lambda^2\right]
\left[\cosh^2(\lambda/2) + n^2\pi^2 x\right]^3}.
\label{function}
\end{equation}
A quick power counting in Eq. (\ref{function}) gives that $f_\alpha(x)$
diverges at $x=0$ (when $\alpha$ is finite). 
A more careful analysis shows that $f_\alpha(x)$
diverges as $x\rightarrow 0$;
but it does so mildly since 
$xf_\alpha(x)\rightarrow 0$.

By inserting Eq. (\ref{drelation}) in Eq. (\ref{edensity})
and recalling that $\kappa'^2=\kappa^2 (1-v^2)$, 
one ends up with
\begin{equation}
\left<{\cal T}^{{\rm T}}{}_{{\rm T}}\right>
=\left<T^{t}{}_{t}\right>-\frac{S^2}{r^6}
f_{\alpha}\left(\kappa'^2/r^2\right).
\label{lia}
\end{equation}
Considering $S\rightarrow\kappa$ in Eq. (\ref{lia}),
$\left<T^{t}{}_{t}\right>$
approaches the ordinary cosmic string expression (and therefore remains finite), 
whereas the term carrying $f_{\alpha}$ diverges 
[one sees that the flaw
mentioned above consists in manipulating improperly the numerator 
in Eq. (\ref{edensity}) when $\kappa'$ is very small].
One might say that $S=\kappa$ plays the role of a 
chronology horizon where mechanisms of chronology protection
are expected to take place. The divergence in  
$\left<{\cal T}^{{\rm T}}{}_{{\rm T}}\right>$
confirms this expectation.
For completeness, the other components of  $\left< {\cal T}^{\mu}{}_{\nu} \right>$
are displayed below
\begin{equation}
\left< {\cal T}^{\mu}{}_{\nu} \right> =
\left(
     \begin{array}{cccc}
\left<T^{t}{}_{t}\right>-(S^2/r^6)f_{\alpha}& 0 & (S/\kappa')\left<T^{Z}{}_{\varphi}\right> & 
(\kappa S/r^6)f_{\alpha}\\
0    &\left<T^{r}{}_{r}\right> & 0  & 0 \\
-(S/\kappa')\left<T^{\varphi}{}_{Z}\right>& 0 & \left<T^{\varphi}{}_{\varphi}\right> & 
(\kappa/\kappa')\left<T^{\varphi}{}_{Z}\right>   \\
-(\kappa S/r^6)f_{\alpha}&  0 & (\kappa/\kappa')\left<T^{Z}{}_{\varphi}\right>& 
\left<T^{Z}{}_{Z}\right>+(S^2/r^6)f_{\alpha}
     \end{array}
\right)
\label{matrix},
\end{equation}
with $f_{\alpha}$ evaluated at $\kappa'^2/r^2$.
The approximate behavior 
of $\left<{\cal T}^{\mu}{}_{\nu}\right>$ as $S\rightarrow\kappa$,
at a given distance $r$ from the defect,
can be obtained from Eq. (\ref{matrix}) by considering the expressions for
$\left<T^{\mu}{}_{\nu}\right>$ in Ref. \cite{lor03}. For example, if $\phi$ is
conformally coupled, it follows that
\begin{equation}
\left< {\cal T}^{\mu}{}_{\nu} \right> =\frac{1}{r^4}
\left(
     \begin{array}{cccc}
-(S^2/r^2)f_{\alpha}      -A &  0 & SB                    & (\kappa S/r^2)f_{\alpha} \\
       0 & -A & 0                    & 0 \\
       -SB/r^2 &  0 & 3A                   & \kappa B/r^2 \\
      -(\kappa S/r^2)f_{\alpha}  &  0 & \kappa B             & (S^2/r^2)f_{\alpha}-A
     \end{array}
\right)
\label{tmunumatrix},
\end{equation}
where $A(\alpha):=(\alpha^{-4}-1)/1440\pi^2$ and $B(\alpha)$ is defined as in Eq. (20)
of Ref. \cite{lor03} [$B(\alpha=1)=1/60\pi^2$]. Clearly the components that diverge as
$S\rightarrow\kappa$ are those containing $f_{\alpha}$.
At this point it is pertinent to note that,
as mentioned previously, 
$\left<\phi ^{2}\right>$ remains well behaved as
$S$ approaches the critical value $\kappa$ [see Eq. (\ref{dphi2})].
If $f_{\alpha}(x)$
were not divergent at $x=0$, $\left< {\cal T}^{\mu}{}_{\nu} \right>$
would also remain finite as $S\rightarrow\kappa$ 
(and that would suggest violation of chronology protection).

The following are simple facts that help to figure out 
the physical implication of the results reported above.
One begins by taking $S$ infinitesimally smaller than $\kappa$,
say $S=\kappa-\delta$. Then, a finite interval of proper time $\Delta\tau$
as measured in the spinning string frame [cf., Eq. (\ref{5le})] would appear an
arbitrarily large interval of time measured in the corresponding
cosmic dislocation frame [cf., Eq. (\ref{dle})], since the latter would be 
traveling nearly at the speed of light (recall that $v=S/\kappa$).
If $\Delta\tau$ is the interval of proper time immediately before the emergence of a 
region containing CTCs (which corresponds to $\delta =0$),
it follows that such an event never would be detected in the 
cosmic dislocation frame.
Moreover, as $\delta\rightarrow 0$, the semiclassical metric tensor
in the cosmic dislocation frame  approaches that found in Ref. \cite{his87}, 
which presents mild contributions from back reaction, as is typically
the case. Considering the divergences in the vacuum expectation value of
the energy-momentum tensor,
such mild contributions (and that is a crucial point) 
would be hugely amplified in the spinning string frame,
suggesting the formation of a ``horizon" when $\delta=0$.
Such a ``horizon" would eventually prevent 
the appearance of CTCs around the spinning string.

\begin{acknowledgments}
The authors are indebted to Renato Klippert for useful discussions.
This work was supported by the 
Brazilian research agencies CNPq and FAPEMIG.

\end{acknowledgments}

\end{document}